\documentclass[twocolumn,twoside,slac]{revtex4}
\usepackage{graphicx}
\usepackage{fancyhdr}
\RequirePackage{xspace}

\def\babar{BaBar\xspace}
\def\mini{Mini\xspace}
\def\pep2{PEP-II\xspace}
\def\BF{$B$ Factory\xspace}
\def\abf {asymmetric \BF}
\mathchardef\Upsilon="7107
\def\Y#1S{\ensuremath{\Upsilon{(#1S)}}\xspace}

\def\FourS {\Y4S\xspace}

\def\invfb   {\ensuremath{\mbox{\,fb}^{-1}}\xspace}
\def\epem       {\ensuremath{e^+e^-}\xspace}
\def\stwob{\ensuremath{\sin\! 2 \beta   }\xspace}
\newcommand{\jprlBase}       {Phys.\ Rev.\ Lett.\xspace}
\newcommand{\jprl}      [1]  {\jprlBase\ {\bf #1}}

\def\svt{{\em Svt}\xspace}
\def\dch{{\em Dch}\xspace}
\def\drc{{\em Drc}\xspace}
\def\emc{{\em Emc}\xspace}
\def\ifr{{\em Ifr}\xspace}

\def\hit{{\em hit}\xspace}

\def\raw{{\em Raw}\xspace}
\def\rec{{\em Rec}\xspace}
\def\esd{{\em Esd}\xspace}
\def\aod{{\em Aod}\xspace}
\def\tag{{\em Tag}\xspace}
\def\hdr{{\em Hdr}\xspace}

\def\Kz    {\ensuremath{K^0}\xspace}

\begin{document}

\title{The \babar \mini}

%

\author{David N. Brown, representing the \babar Computing Group}
\affiliation{Lawrence Berkeley National Lab, USA}

\begin{abstract}
\babar has recently deployed a new event data format referred to as the \mini.
The \mini uses efficient packing and aggressive noise suppression to represent the average
reconstructed \babar event in under 7 KBytes. The \mini packs detector
information into simple transient data objects, which are then aggregated into
roughly 10 composite persistent objects per event.  The \mini currently uses Objectivity
persistence, and it is being ported to use Root persistence.  The \mini contains enough
information to support detailed detector studies, while remaining small and fast enough
to be used directly in physics analysis.  \mini output is customizable,
allowing users to both truncate unnecessary content or add content, depending on their
needs. The \mini has now replaced three older formats as the primary output of \babar
event reconstruction.  A reduced form of the \mini will soon replace
the physics analysis format as well, giving \babar a single, flexible
event data format covering all its needs.

\end{abstract}

\maketitle

\thispagestyle{fancy}


\section{The \babar Experiment}

\babar is a multi-purpose detector operating at the
\pep2 \abf.  \babar has been taking data at and
near the \FourS resonance since 1999,
and has accumulated roughly 110 \invfb of luminosity to date.  \babar is fairly typical of
modern High Energy Physics apparati, consisting of several quasi-independent detector
subsystems arranged roughly concentricly about the \epem interaction point.
The innermost subsystem is the Silicon Vertex Tracker (\svt), with roughly
150K readout
channels.  Outside of the \svt is the Drift Chamber (\dch), with 
roughly 7K readout
channels.  Outside the \dch is the Cherenkov Detector (\drc), with 
roughly 11K readout channels.
Outside the \drc is an Electromagnetic Calorimeter (\emc), 
with roughly 7K readout channels.
Outside the \emc is the Instrumented Flux Return (\ifr), 
with roughly 60K readout channels.

\babar was an early adopter of C++ and OO programming in HEP, and the vast majority of
our software is written in C++ \cite{chep_jacobsen}.
\babar has used Objectivity as the
primary technology for storing event data \cite{chep_quarrie}, however we are planning to change
to a Root based event store by the end of 2003 \cite{chep_kirkby}.

\section{\babar Event Data Format History}
\label{sec:formats}

\babar's original software design \cite{eventdesign} 
proposed several complimentary
event data formats, as described in table~\ref{tab:formats}.
These formats were intended to satisfy different use cases, from
quality control to reconstruction to calibration to physics analysis,
with each format optimized for some specific purposes.  
Each format was written to separate Objectivity databases (files),
so that they could be accessed and managed independently.
The \raw format is an
Objectivity transcription of the raw data readout by the detector
online system, and was intended to be used as the input
to the reconstruction chain.
The \rec format represents the reconstructed physics objects,
and was intended to
be used for detector studies, for detailed analysis, 
and for single event display.  The \esd format is
a summary of the reconstruction results, and was intended to be
the primary format used for high-statistics physics analysis.
The \aod format was intended to store highly processed
information specific to physics analysis.
The \tag format was intended to store booleans to index
and quickly select events.  The (\hdr) format allows
events to `borrow' some subsystem data from other events, and was
intended to support partial re-processing of individual subsystems.

By 2001, these data formats and their usage in \babar had stabilized.
As shown in table~\ref{tab:formats}, many of the data formats were
not actually used.  Additionally, the \aod and \tag formats
were considerably larger than foreseen, and had taken on different roles than
originally intended.  Subsequent sections explain
why the formats were not used according to the original design,
and how that led to the development of the \mini.

\begin{table}[t]
\begin{center}
\begin{tabular}{|l|c|c|c|}
\hline
Format & Design Size & Actual Size & Usage \\
\hline
\hline
{\em \raw} & 25 KBytes & 50 KBytes & Unused \\
\hline
{\em \rec} & 100 KBytes & 120 KBytes & Unused \\
\hline
{\em \esd} & 10 KBytes & 7 KBytes & Unused \\
\hline
{\em \aod} & 1 KBytes & 3 KBytes & Analysis \\
\hline
{\em \tag} & 100 Bytes & 1 KByte & Selection \\
\hline
{\em \hdr} & 0 & 4 KBytes & Navigation \\
\hline
\end{tabular}
\label{tab:formats}
\end{center}
\caption{\babar Objectivity event data formats circa 2001.  \raw refers
to raw data, \rec to reconstructed data, \esd to event summary data,
\aod to analysis data, \tag to event selection data, and \hdr to
event header data.}
\end{table}

\subsection{The \babar Persistence Design}
\label{sec:persistent_design}

\babar's original persistence design can be summarized as translating transient
objects and transient object relationships into equivalent
persistent objects and persistent references, as illustrated in
figure~\ref{fig:trans2persist} for the specific case of reconstructed
tracks.  The persistent objects were clustered into the various databases
according to how it was anticipated they would be used.  This design
established the now-standard 
$ transient \rightarrow persistent \rightarrow transient$
paradigm in a straightforward way.  This design
allowed analysis jobs running on
(\esd) data to retrieve reconstruction details about objects
on demand, by following a link back into the \rec database.  
This was considered an important example of how an
OO database event store might provide significant new
functionality compared to sequentially organized
data storage technologies.

\begin{figure*}
\centering
\includegraphics[width=145mm]{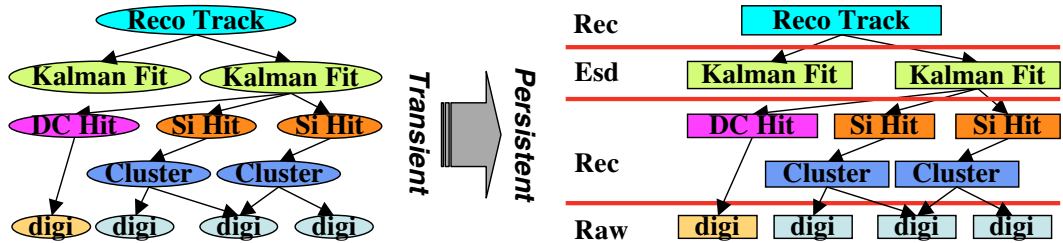}
\caption{Diagram of the object tree representing a reconstructed
track in \babar, on the left in transient form, and on the right
converted to persistent objects according to \babar's original
persistence model.  The persistent objects are grouped
according to which database they were stored in (the database labels are
explained in section \ref{sec:formats}).} \label{fig:trans2persist}
\end{figure*}

The literal translation of complex transient object trees
to persistent object trees resulted in a fragmented structure, where
different parts of a single physics object 
(a track in figure \ref{fig:trans2persist}) were distributed
across several databases.  This effectively coupled
the data formats and database files.
For instance, a job reading tracks from \esd depended on the
\rec database to provide the top level tracking persistent object.
This coupling added enormously to the IO burden and the disk footprint of an analysis
job running on \esd.

Similarly, the \rec format design required that
transient objects be rebuilt from their constituent \raw data.  This coupled the \rec
format to the \raw, and required that a job reading \rec data
pull in essentially the entire reconstruction code base.
A \rec job thus consumed a similar amount of
resources (cpu, memory, and disk) as the original reconstruction job.

An additional difficulty to accessing the \rec format was that
the large size of the \rec databases precluded storing them on disk.  Instead, they
were accessible only through staging.
As the staging space at SLAC was originally very limited,
dynamic staging through the Objectivity HPSS interface was disabled,
forcing users to stage \raw and \rec databases by hand.  This
tedious and error-prone operation proved impractical for the vast
majority of \babar physicists.

\babar originally considered having the online
system write raw event data directly in
Objectivity \raw format.  However, since OO database technology
was new and relatively untested, a more conservative approach
was taken, where the online system writes a flat file version
of the raw data, which can then be transcribed into Objectivity
\raw format.
It was then found to be more efficient to
reconstruct events by directly reading the online raw data.
The \raw format was thus recast as an output instead of an input of
reconstruction.

The \raw format was used to pass data between the \babar simulation executable
and its reconstruction executable.  In 2002 \babar developed a
monolithic simulation plus reconstruction executable, which
eliminated the need for \raw as intermediate storage.  This monolithic
simulation executable has been used for official \babar Monte Carlo generation since early
2003.

\subsection{The Unused Formats}

The poor performance of jobs reading the
\rec format, coupled with the difficulty
of accessing \rec and \raw data,
made them nearly impossible to use for
analysis or detector studies.  Only a handful
of physicists on \babar ever made use of either of
these data formats, and those uses
typically involved very small event samples.
Instead, most calibration and monitoring tasks
run on raw online data, invoking reconstruction code
as necessary to build objects, or on the expanded 
\aod format.

The \esd format was not finished in time for \babar's first data,
partly because its development was given
lower priority compared to developing the \raw and \rec formats.
It was also felt that some experience with the \babar detector
and data analysis was necessary before
the \esd could be correctly designed to meet the
needs of experiment.  It was instead foreseen that \babar's first data
would be analyzed using the \rec format, from which experience
the \esd format could be completed.
Since \babar never used \rec to analyze first
data (see section \ref{sec:evolution}), the \esd format was never completed
and never used.

The capability of events to borrow content from other events
via the \hdr format was also never used, mostly because
\babar never encountered a situation which required
partial re-processing.  The problem of managing the
dependent event databases this would have created
was also never addressed.  The large size of the
\hdr format was principly due to large string arrays describing
component names.  The \hdr was
recently redesigned, greatly reducing its size
(see \cite{chep_yemi}).

\subsection{The Evolution of \aod and \tag}
\label{sec:evolution}

Since it was imperative that \babar start developing its
analysis procedures even before first data, 
and since the formats originally intended to be used for analysis were
effectively unusable,
\babar decided to expand the \aod format so
that it could be used for physics analysis.  This was implemented
by including a `four-vector' summary of the reconstruction results,
together with `quality values' to describe some detector-specific
details.

The \aod format was spectacularly successful in enabling
analysis of \babar's first data, allowing important physics results
to be produced in a timely way.  The \aod
has since been \babar's primary physics analysis format, and
it underlies all the physics results published to date.  This format does
however have several limitations.  For one, the only track fit result
stored in \aod assumes the particle which generated the track
was a $\pi$ ( \babar reconstruction provides all (5) stable particle
track fit results).   Because the energy loss due to
material interactions is mass-dependent, this limitation introduces a small
momentum bias when the particle is not a $\pi$.

Another limitation of \aod is that it 
reduces the detector information to `a set of
numbers' extracted from reconstruction objects.
This greatly reduces the benefits which
\babar might have been obtained from
from using OO design and interfaces in
analysis.  It also effectively
isolated \babar analysis from reconstruction, making it impossible
to port code developments from one side to the other.

The \aod format also provides a rigid persistent representation of an event,
with no way to tailor its contents for specific
use cases.  Consequently, most \babar analyses operate by dumping the
\aod format into an ntuple, adding data content according
to thier needs.  This effectively doubles \babar's analysis data storage needs.
It also decouples analyses from each other, as different analysis working
groups have developed different ntuple representations of the \aod format.
The redundency and inefficiency of this analysis methodolgy was a strong
motivation for \babar's new computing model, described in section \ref{sec:new_model}.

The \tag format also evolved when confronted with first data.
To provide more flexibility when selecting events,
the \tag format was expanded to include
floating point and integer values as well as booleans.  Thus the intent
of the data formats was `pushed down' one level compared with the
original design, with the \aod taking
the role intended originally for \esd, and \tag taking the role
intended for \aod.

By contrast with \raw, \rec, and \esd, the \aod and \tag formats were developed
to be completely independent of the reconstruction objects.  This avoided the
interdependency problems of the unused formats, at the cost of
allowing no way to navigate between physics analysis
objects and the reconstructed and/or raw data from which they were derived.

\subsection{The Data Format Gap}

Because of the evolution of \babar's data formats,
a large gap had developed,
with no practical way to access information between raw online
data and physics 4-vectors.  This gap made performing routine functions
like calibrations and detector diagnostics
difficult and time consuming.  The gap also severely limited the ability
to study detector effects in physics analysis.  This gap
also prevented \babar from developing a usable single event display.
The data format gap was first officially recognized in 2000
in the report of an internal review of \babar computing \cite{cmwg1}.

The \svt provides one example of how the data format gap caused problems.
In order to obtain optimal tracking resolution, the
positions of the \svt wafers ({\em alignment}) must be derived from the data.
The \svt alignment procedure needs both low-level data (hits) and
high-level data (tracks) to perform this task.  Because of the data format gap,
it was found that the most efficient way of doing this was to read raw data,
and reconstruct the tracks in the alignment job.
The alignment job was therefore very slow, and
the total procedure had a turnaround time of roughly 1 month.
This was found to be longer than the time interval over which the \svt
wafer positions were stable.  The long turnaround time also
stifled development of the alignment procedure,
as it took too long to test changes.  The net
result was that the \babar used a poor alignment of the \svt
in reconstruct early data, degrading the effective resolution
of the detector,
and introducing sizeable systematic errors in many physics analysis.
\svt misalignment caused the dominant systematic error in \babar's first
\stwob publication \cite{sin2beta}.

\subsection{The Origins of the \mini}
\label{sec:origins}

To solve the \svt alignment problem, a new
data format was developed for storing tracks.
This new format stored tracks together with their
hit data in a compact, flexible, and efficient structure.
This new track persistence format was
used to design a new \svt alignment procedure, which reduced the turnaround time
to roughly 1 day, and
which produced measurably better physics results.
The success of the new track data format and the new alignment procedure
inspired a larger effort to develop a new data format for all of \babar
based on the same basic design.  This new format was referred to as the \mini.

The \mini development project was officially
begun in early 2001.  A prototype of the \mini was produced in a
complete re-processing of the \babar data sample started in 2001.
The prototype \mini was used to perform many detector studies, and to
refine the \mini design.
Unfortunately the \mini prototype did not include any
information from the \emc, and so was not usable for physics analysis.
The first complete version of the \mini was released
in early 2002.  The design and implementation of the \mini is described
in the remaining sections.

%


\section{The \babar \mini Design Goals }

The main design goal
of the \mini was to persist the results of event reconstruction.
To avoid the problems of the \rec format (which had the same
goal), the \mini was also required to be small
($<$ 10 KBytes per event),
self-contained (no references to objects outside the \mini),
and fast (support reading at roughly 20 Hz, equivalent to
reading the \aod format).

Another goal of the \mini design was
to provide access to sufficient detector detail to support 
standard calibration, alignment, diagnostics, and algorithm development.
This capability would also allow the \mini to support a detailed
single event display. 

The \mini was required to be able
to follow changes in {\em Conditions}
(alignment and calibration parameters), so that
users could benefit from improved parameters without having to
wait for data re-processing.  This requirement would also
allow analysis users to easily propagate calibration and alignment
uncertainties to systematic errors in their analysis,
by simply re-running with altered parameters.

To maintain compatibility with reconstruction, The \mini
was required to provide access
through the interfaces of actual reconstruction classes,
without any significant loss in accuracy or content compared to
the original reconstruction results.

To support specialized use cases, the \mini was designed
to allow users to customize the persistent content according
to their needs.  This feature is relied on heavily in \babar's
new computing model, as a way of reducing the need to dump
data into ntuples in order to do analysis.

To leverage \babar's huge physics analysis code base, the \mini
was required to be compatible with
the existing analysis framework.  Explicitly, the goal was that
an average user be able to convert their analysis job
to read the \mini instead of \aod without changing
any physics-related code, and that the results obtained
be equivalent (within floating point precision) to those obtained
with the original \aod format job.

\section{ Implementation of the \mini}

The \mini design goals
require both access to the full detector detail
through reconstruction interfaces, together with full compatibility and
similar performance as the \aod (4-vector summary) format
for physics analysis.  The \mini satisfies these contradictory requirements
by storing both high-level objects
(tracks, calorimeter clusters, Cherenkov rings, particle ID, etc),
and the low-level objects
(\dch hits, \emc crystals, \drc phototubes, etc) from which they were made.
This results in some redundancy, as some content of the high-level
objects can also be extracted from their constituent
low-level objects.  Redundancy is generally considered a bad idea
in data storage, as it can cause consistency problems.  It was accepted
for the \mini design as it afforded a large
(~factor of 10) performance improvement when reading high-level objects
(see \ref{sec:performance} for details), and because the \mini
access mechanism includes safeguards against
inconsistent data usage (see section \ref{sec:access} for details).

\subsection{ High-level objects in the \mini}

High-level objects in the \mini store the set of references to the low-level
objects from which they were built, thus preserving the essential
information of the pattern-recognition algorithm.
High-level objects also store 
references to the  other high-level objects they depend on.
For instance, \drc rings store a list of
\drc \hit references, plus a reference to the track used to seed
and fit the ring.  

High-level objects also store the results of
cpu-intensive functions which their transient class
supports.  For instance, track objects store the results of
running the Kalman filter fit.  Because these functions use
the associated low-level objects, these stored results are
redundant with the low-level objects themselves.
Because these functions
were invoked during reconstruction, stored results
implicitly depend on the {\em Conditions}
which were used when reconstruction was run.  Thus, stored results
of high-level objects do not follow changing {\em Conditions}.
To follow new {\em Conditions}, the stored results cannot be used, and
the original functions must be called on rebuilt
transient objects.
Details of how the \mini
can be configured to use (or not) stored function results
is described in section \ref{sec:access}.

\begin{figure*}
\centering
\includegraphics[width=145mm,angle=-90]{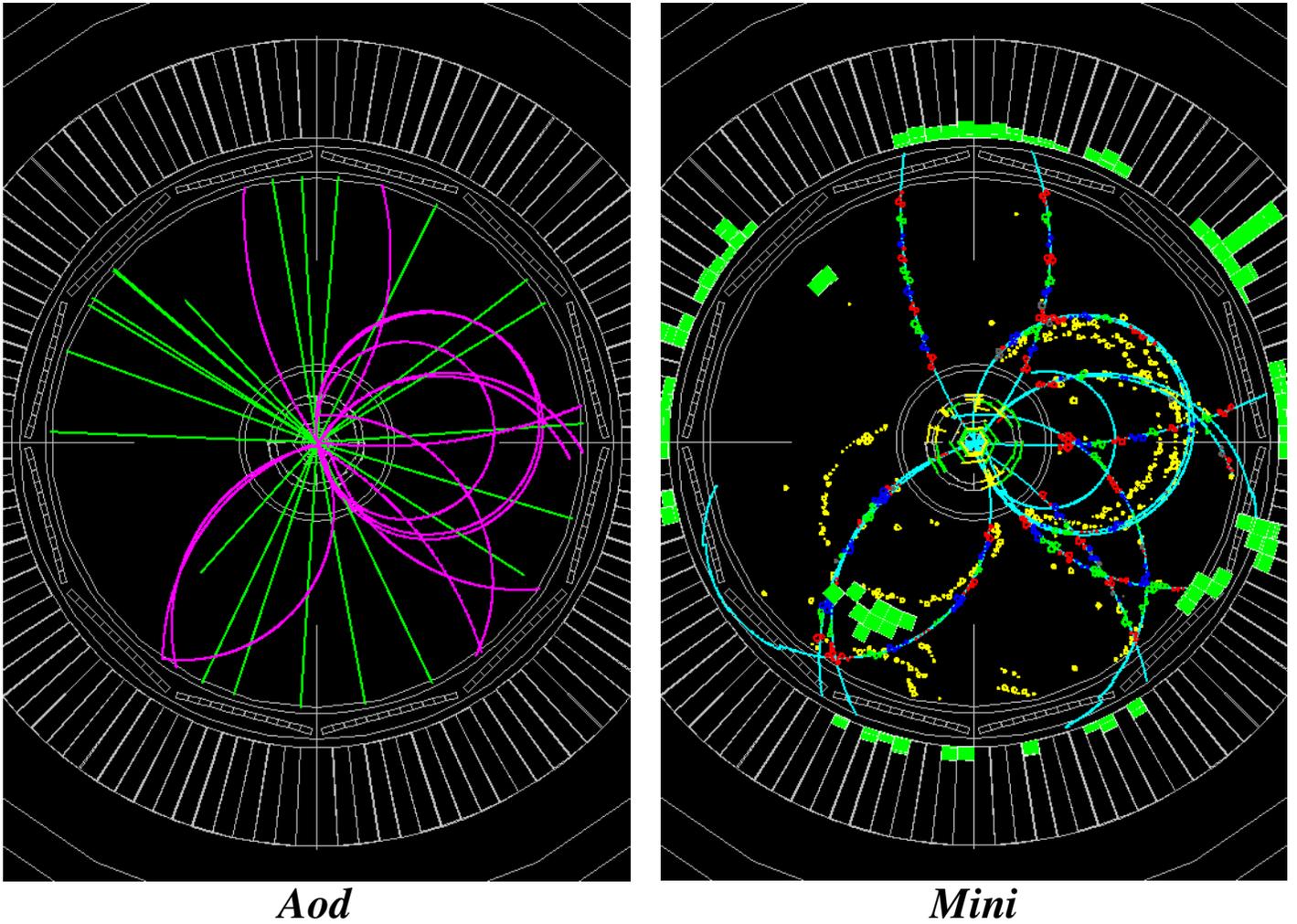}
\caption{Single event display of a typical \babar multi-hadron event in \aod format
on the left, and \mini format on the right.  In \aod, tracks are modeled
as perfect helices, and neutral objects as 4-vectors.} \label{fig:evtdisplay}
\end{figure*}

\subsection{Low-level objects on the \mini}

Where possible, the low-level objects in the \mini store
raw detector readout information instead of physical quantities.  Thus
\dch hits are stored as TDC values and wire numbers instead
of physical times and positions.
Physical quantities are then computed from the raw data on the
\mini using conversion algorithms and
{\em Conditions} data, as implemented in the transient
low-level reconstruction class accessor functions.
This allows the \mini to follow {\em Conditions} changes,
and to provide consistent results with reconstruction.

For some subsystems,
the raw detector data are very large, and they must be
compressed before being stored on the \mini.  In these cases, the 
compressed information is still stored in detector units.
For instance, \svt hits are compressed to store
the average cluster position instead of all the individual strips in a
cluster, but the average position is expressed in strip coordinates.

The \mini also stores a subset of
low-level objects {\em not} associated
to any high-level object.
Monte Carlo and other studies have shown that
many of the unassociated low-level objects
were generated by particles produced directly or indirectly
in the \epem collision.  Unassociated low-level 
objects can be used to identify physics objects missed
due to reconstruction inefficiency, or to search for unusual physics signals not
found by the standard reconstruction.
Associated and unassociated low-level objects can also be combined
to create a `complete' set of low-level objects.  This allows
the \mini to be used
to develop and test pattern recognition algorithms, or to be used
as a source for partial re-processing.

Unfortunately, most unassociated low-level objects in
a typical \babar event do not come from the \epem collision.
Storing all of them would therefore bloat the \mini and degrade its contents.
Instead, only unassociated low-level objects which pass
stringent quality cuts are stored on the \mini.
For instance, only those unassociated \svt hits
whose arrival time is consistent with the reconstructed
event time are stored on the \mini.  This cut reduces the number
of unassociated \svt hits by roughly a factor
of 20, while keeping roughly 90\% of the `real' unassociated
\svt hits.

\section {\mini Persistence}
\label{sec:persistence}

The \mini was first implemented using Objectivity persistence, and
it has recently been ported to use Root persistence, as part
of \babar's new computing model (see section \ref{sec:new_model}).
A large part of the \mini's success was due to strict
adherence to a few basic persistence design principles, described
in the following sections.

The \mini persistence is controlled by the standard \babar persistence
mechanism.  A dedicated {\em loader} module is run for each detector
subsystem, which creates the {\em scribes} responsible for
translating specific transient objects into their persistent counterpart.
The event key by which a scribe identifies its transient object is configurable
through the loader module Tcl interface.
Thus the user can control the content of the \mini by choosing
which loader modules to run, which scribes to create,
and which transient objects the scribes should convert,
configurable through control scripts on standard executables.

The configurability of \mini persistence 
was used to improve the efficiency of the
\svt alignment procedure (see section \ref{sec:origins}).
By reading a custom reduced \mini holding just selected \svt
track information, the iterative part of the alignment procedure
was sped up from several hours to just 10 minutes per iteration.
As convergence required hundreds of iterations,
this speedup was essential for producing the
23 different \svt alignment sets used in the 2002 data re-processing.

For technical reasons, the \mini was not placed in a new database.
Instead, the \esd database was cleared of all previous content, and the
\mini was placed there.  The \mini thus completely replaced the
original \esd format.

\begin{figure*}
\centering
\includegraphics[width=145mm]{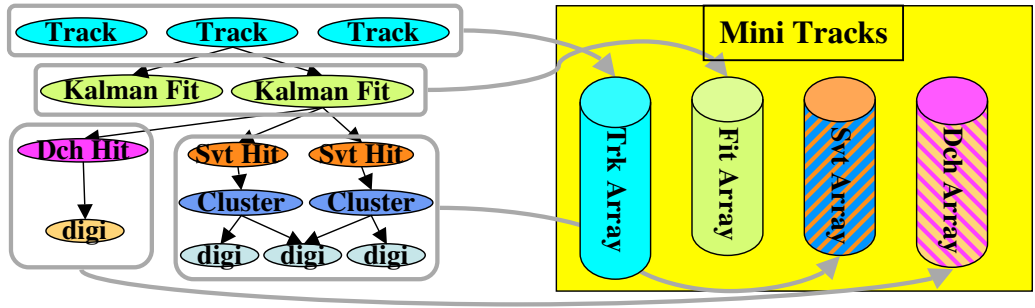}
\caption{Diagram of the object tree representing a reconstructed
track in \babar, on the left in transient form, and on the right
converted to \mini persistent format.
The \mini stores all tracks in an event
in a single persistent object.  Track-specific details are described by
the embedded objects stored in the different persistent arrays.}
\label{fig:minipersist}
\end{figure*}

\subsection {The Persistent Composite Design}

The \mini persistence design is based on
the {\em persistent composite} design pattern,
in which a single persistent object holds the data
for a collection of transient objects of a given type.
Using this design, the \mini stores the
contents of a \babar event in just 11 persistent objects,
minimizing the impact of the 12 bytes per persistent object
Objectivity overhead.  A graphical representation of the \mini persistence design
is shown in figure \ref{fig:minipersist}.

In the persistent composite design pattern, the
contents of a transient object collection are
stored in persistent arrays (ooVArrays for Objectivity)
of {\em embedded} objects, which have a one~$\leftrightarrow$~one
relationship with transient objects.

Embedded classes translate to and from their
transient counterparts, but otherwise provide no interface.
They are implemented as simple structs of primitive data types,
with no dependence on any persistence technology.
Because they are persistence-free, the same embedded classes
can be used by different persistence mechanisms,
making it easy to port persistent composite classes
other persistence technologies.

Associations between objects on the \mini are stored
in the persistent composite objects as 
a single reference (OORef for Objectivity) to other
persistent composites.  Specific objects in other
composites are then referenced as the {\em index} into the
corresponding embedded array.  
This results in much less
overhead than storing explicit references, as an
index (typically 2 bytes) is much smaller than an OORef (12 bytes).

\subsection{Data Packing on the \mini}
\label{sec:packing}

To minimize the size of the \mini, its data contents are
packed, according to the following rules.

\begin{itemize}
 \item{Boolean data are stored as a single bit.}
 \item{Integer data are stored using as many bits as required by their range.}
 \item{Float data are packed and stored as integers.  The 
        {\em Least Significant Bit } of the packed data (LSB) corresponds
        to roughly 1\% of the intrinsic detector resolution
        of the quantity being stored.
	Float data with an extreme natural range are packed logarithmically,
	using an algorithm which is locally flat
        to avoid binning artifacts in histograms (see figure \ref{fig:packing})}.
\item{Packed integer and float values are truncated at
      physically reasonable ranges, not `worst possible' ranges.
      Values beyond the physically reasonable range are flagged as under or overflows.}
\item{Strings are stored as a key (integer)
      in a string~$\leftrightarrow$~integer map.  The
      map is stored outside the \mini event data.}
\item{Small data fields are combined (bitwise OR)
      to fill a standard type (char, short, or long) word.}
\item{Data members of embedded classes
      are all aligned to either char, short, or long word boundaries (one choice per class),
      to ensure that embedded object arrays are compact in memory.}
\item{To avoid the Objectivity overhead of storing and retrieving virtual tables,
      embedded classes have no virtual functions, {\em including no virtual destructor}.}
\item{Direct data members of persistent composite classes are aligned to long word boundaries,
	to be consistent with Objectivity persistent object alignment}
\item{To avoid creating persistent memory fragments, variable arrays (ooVArrays)
      are sized exactly once, either on initialization or in the constructor body.}
\end{itemize}

\section{Accessing \mini Data} 
\label{sec:access}

\begin{figure*}
\centering
\includegraphics[width=80mm]{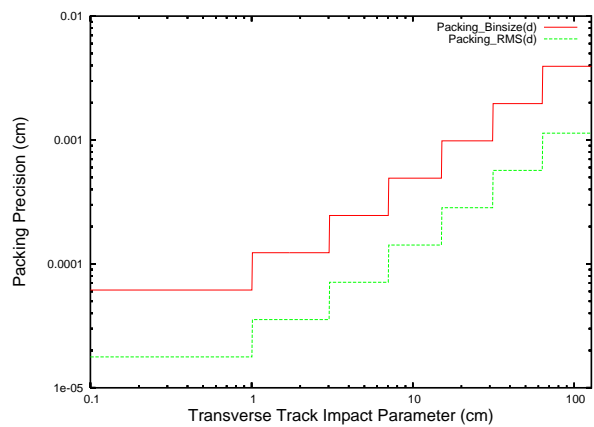}
\caption{Resolution of the globally-logarithmic, locally flat packing
algorithm used in the \mini, as applied to packing the track
transverse impact parameter.  This quantity has a
detector resolution of roughly 10 microns, and range of values from 0 to 80 cm.
The packing algorithm employed is extremely efficient to unpack, involving only
2 floating point operations (one addition and one multiplication).}
\label{fig:packing}
\end{figure*}

As described in the previous section,
the \mini persists several levels of data which
overlap in content.  The \mini is designed so that
the user must decide which
level of detail is appropriate
when reading back the \mini.
In making this decision, the user must balance the 
greater detail and (potentially) greater accuracty
which are available when reading
low-level data, against the greater speed possible
when reading high-level data.
The \mini persistence provides a very precise degree
of access control, so that some objects may be read with 
high precision while others are read with lower precision.
Similarly, it is possible to read an event initially
at low precision, and later upgrade some or all objects to
higher precision once the event or the objects in it
pass cuts.  Maintaining coherent and correct data contents
under these general conditions is however difficult, and
involves a level of expertise beyond that of the average
user.

To make it easier for users to correctly configure reading the \mini,
a set of self-consistent access
modes are provided which roughly span the available
options.  While some users may need to optimize their
\mini jobs by directly controling the persistent access,
it is expected that most
\mini users will choose one of these five access
modes:
{\em micro}, {\em cache}, 
{\em extend}, 
{\em refit}, and {\em raw}.
The access mode is set in a user job through
the {\em levelOfDetail} global Tcl variable, which is then used
by the sequences which read and prepare the \mini
data for analysis.  The specifics of these different modes
are described below.  A comparison of the access speed
in these different modes is given in table \ref{tab:modes}.

The \mini user can also control how
the \babar {\em Conditions Database} \cite{chep_babar}
is accessed.
For instance, a user can configure their \mini job
to use the same {\em Conditions} as
were used when the data were originally processed, or the most recent
{\em Conditions}, or even to override the 
{\em Conditions Database} and use
explicitly-provided constants.  {\em Conditions} access configuration is most
relevant to refit and raw modes.

\subsection{Cache Mode}

Cache mode refers to reading the \mini so
that high-level objects are built from the
stored function results instead of low-level data.
Only a summary of the low-level information 
can be obtained in cache mode.  For instance,
in cache mode, the track transient object  can provide
the number and logical identity of the hits from which it was fit,
but it cannot produce the actual hit objects.
Since a cache mode job doesn't read or process any low-level data,
it is much faster than a refit job run on the same data.

In cache mode, all the stored track fits can be used.
The default version of the mini stores all the
unique mass hypothesis Kalman fits, evaluated at their point
of closest approach to the Z axis, plus the
$\pi$ fit evaluated where the track exits the tracking volume.

\subsection{Micro Mode}

Micro mode is a variant of cache mode where
some features are turned off, in order to make the \mini behave
more like the \aod format.
For instance, since \aod stores only $\pi$ track fits, 
in micro mode the \mini track provides only the $\pi$ fit.
Micro mode is intended to make it easy to compare and validate
the \mini against the \aod format.
Since micro mode is no faster than cache mode,
and yet returns less accurate values, it is not recommended
for use in physics analysis.

\subsection{Extend Mode}

Extend mode is a variant of cache mode in which
the validity range of a track is extended from
the range of the fit result stored on the \mini, up
to the first hit.
Otherwise, extend mode behaves exactly as cache mode.
The persistent data read in extend mode are exactly the same as
in cache mode, but since more tracking functions are called,
extend is somewhat slower to read than cache.

In extend mode, the fit results stored on the \mini
are interpreted as a multi-dimensional `hit',
constraining all the track parameters to the
stored fit values.  These `constraint hits'
are used to create a Kalman track fit object,
which is an instance of the same Kalman fit class
used to fit tracks in \babar reconstruction.
The Kalman fit adds the effects of passive material and magnetic
field distortions as the track traverses
the the detector, extending the 
range over which the stored track can provide physically accurate
parameters.  Since hits are not read in extend mode,
extended tracks are valid from the origin out to the first hit.

An example where extend mode is useful is
reconstructing long-lived particles which decay outside
the beampipe, such as \Kz.  In cache mode (and when using
the \aod format), these particles are vertexed using fit results
measured inside the beampipe.  In extend mode, track fit results are
measured at the decay vertex, so that the
reconstructed parameters of the \Kz
are more accurate and less biased.

\subsection{Refit Mode}

In refit mode, the function results stored with the
high-level objects are ignored, and high-level transient
objects are rebuilt
from constituent low-level objects.
Because refit mode involves reading more data and
performing more computation to create the high-level objects,
it is substantially slower than either cache or extend mode.
Because new {\em Conditions}
are read and used when rebuilding the transient objects,
a refit mode job can follow changing {\em Conditions},
or even changes in some reconstruction algorithms.
Refit mode is intended to support detector studies,
single event display, specialized analyses that depend on low-level
data, and analyses that need to use new {\em Conditions}
or algorithms.

\subsection{Raw Mode}

In raw mode, high-level objects stored on the \mini are ignored.
Both assigned and unassigned low-level objects are
are read and combined together into `complete' lists, and the
reconstruction pattern recognition algorithms are run on those.  Raw
mode is intended to support development of
pattern-recognition reconstruction algorithms, to support
re-processing, and to support event mixing studies.
Because raw mode invokes
pattern recognition algorithms,
it is slower than refit mode.

While the high-level
objects created when reading the \mini in raw mode are similar to those
read in the other modes, they are not necessarily identical,
as the initial sets of low-level objects are not
exactly the same as those used when running reconstruction on raw online data.

Raw mode is still under development as a user option, though 
it has been tested in a limited form.

\section{Performance of the \mini}\label{sec:performance}

General performance numbers for the \mini, such
as size on disk and read speed under various conditions are
listed in tables \ref{tab:modes} and \ref{tab:evtsizes}.
Performance of the \aod format is given for comparison.
As efforts to optimize the read speed of the \mini have only just begun,
these numbers should be considered provisional.
Table \ref{tab:time} gives a breakdown of where time is
currently spent
in a typical \mini analysis job.  This clearly shows that unpacking
data plays a very minor role in the performance.

\begin{table}
\begin{center}
\begin{tabular}{|c|c|c|}
  \hline
Data  & Generic & Multi-hadron \\
  \hline \hline
  \mini & 6.4 KBytes & 10.0 KBytes \\
 \hline
 analysis reduced \mini & 1.8 KBytes & 3.2 KBytes \\
  \hline
  \aod  & 1.8 KBytes &  2.7 KBytes  \\
  \hline
\end{tabular}
\end{center}
\caption{The average (compressed) size of \babar events stored
in \mini, analysis reduced \mini and \aod formats.  Results
for the analysis reduced \mini are based on a prototype.}
\label{tab:evtsizes}
\end{table}

\begin{table}
\begin{center}
\begin{tabular}{|c|c|c|}
\hline
mode & Generic & Multi Hadron \\
\hline
\hline
micro & 45 Hz & 24 Hz \\
\hline
cache & 45 Hz & 22 Hz \\
\hline
extend & 28 Hz & 14 Hz \\
\hline
refit & 5.3 Hz & 2.4 Hz \\
\hline
raw & 3.3 Hz *  & 1.0 Hz*  \\
\hline
\aod & 246 Hz &  173 Hz  \\
\hline
analysis reduced \mini & 96 Hz &  -  \\
\hline
\end{tabular}
\end{center}
\caption{The event rate reading the \mini
in different modes on an 1.4 GHz Pentium III Linux machine.
The times for raw mode were estimated using the \babar reconstruction
executable, as this \mini mode has not yet been fully implemented.  Results
for the analysis reduced \mini are based on a prototype.}
\label{tab:modes}
\end{table}

\begin{table}
\begin{center}
\begin{tabular}{|c|c|}
\hline
Operation  & \% time \\
\hline \hline
Reconstruction transient creation + deletion  & 35 \\
\hline
Objectivity data read & 25 \\
\hline
Physics interface adapter & 20 \\
\hline
Event loop overhead  & 10 \\
\hline
Data unpacking & 0.1 \\
\hline
\end{tabular}
\end{center}
\caption{The fraction of time spent in various operations
when reading the \mini in cache mode in a standard physics analysis job.}
\label{tab:time}
\end{table}

\section{\babar's New Computing Model}
\label{sec:new_model}

In April 2002 \babar computing was reviewed by a combined internal
and external review board.  Among other recommendations,
the report of this committee \cite{2002_review}
suggested that \babar reconsider its Analysis Model in light of
the opportunities offered by the \mini.  In response to these
recommendations, a new Computing Model
was adopted by the collaboration in December 2002.  This model introduces
two major changes, first that the \babar event store be converted to use
Root persistence instead of Objectivity, and second that the existing physics
analysis format (\aod) be replaced with a new format more consistent with the \mini.
After some discussion, a reduced \mini customized for analysis has been chosen
as the \aod format replacement.

To replace the \aod format, the analysis reduced \mini must have a similar
size on disk and read-back speed as \aod.  The starting point for
achieving this is to store only those quantities referenced
in cache mode.  Performance results from a prototype analysis reduced \mini
are given in tables \ref{tab:modes} and \ref{tab:evtsizes}, showing that
it is similiar to \aod.  A major effort is now underway at \babar
to improve the read-back performance.  Based on profiles of a standard
analysis job, the largest time sinks come from
inefficiency in the reconstruction code invoked when reading the \mini, and in
the analysis interface to the \mini
(see table \ref{tab:time}).  Based on
the problems already identified, the read speed is expected in
increase by between a factor of 2 and 10.

In the new \babar Computing Model,
the analysis reduced \mini will store only the cache mode information.
The remainder of the \mini information will be stored in a separate
file.  The complete \mini will be accessed by reading both
the reduced and remainder information.  Thus \babar will store event
data in a coherent format, split into pieces specialized for analysis
and detector studies, with no redundancy and easy navigation between the
two pieces.

A requirement of the new computing model is that the \aod replacement
be accessible interactively.  As part of satisfying this requirement,
\babar has chosen Root as the
persistence technology for the \aod replacement,
since it has been shown to work as a HEP
event store technology both at \babar and elsewhere, and because the
Root/CINT interface is a standard interactive access mechanism.  To
satsify this requirement, the \mini is being ported to Root persistence.
The \mini Root persistent implementation uses
base classes developed at \babar which allow interactive
access to packed data contents of embedded
objects, by dynmaically linking class functions
into Root \cite{echarles}.

A key feature of the new computing model is the ability 
to create custom output
streams for physics groups, by exploiting the configurability of the
\mini.  Coupled with the interactive access capabilty afforded by
Root persistence, it is hoped that custom streams can replace the
\aod format dump ntuples used in most analyses.  This will substantially
reduced the computing and human resources used in analysis.

Because \babar is a functioning experiment, the new computing model
must be introduced in a way that does not disrupt ongoing efforts,
and quickly enough that its benefits can be exploited before the
experiment ends.  The plan is to develop and
deploy the new computing model within calendar year 2003.

\section{Conclusions}
\label{sec:conclusions}

\babar has introduced a new event data format refered to as the
\mini.  This format addresses deficiencies in \babar's older formats.

\babar has just completed a full data re-processing, in which the
complete \mini replaced the unused \raw, \rec, and \esd formats.
This reduced the volume of data produced in reconstruction
by a roughly factor of 10, significantly improving the efficiency of the event
reconstruction farm, and requiring half as many
data servers compared to previous processings.
\babar's data storage costs were also cut by roughly the same factor of 10.

\babar is now starting to use the \mini for physics analysis.
An ambitious new computing model has been adopted, in which
a reduced form of the \mini will replace the current physics analysis format.
When the new computing model is deployed in late 2003,
\babar will have a coherent event data format covering most
of the needs of the experiment, finally satisfying the intent of the
original format design.


\begin{acknowledgments}
The authors wish to recognize the achievements of the
\babar offline software developers who designed
and implemented the \babar reconstruction framework
and event persistence, which has been spectacularly
successful in enabling \babar to produce high quality
physics results, and which laid the foundation for 
developing the \mini.
\end{acknowledgments}


\end{document}
%